\documentclass[a4paper,11pt]{article}
\usepackage{jinstpub} 
\usepackage{lineno}
\usepackage{bbm}
\usepackage{caption}
\usepackage{subcaption}
\usepackage{comment,mfirstuc}
\usepackage{ulem}

\newcommand{\addComment}[2]{
  \expandafter\newcommand\csname #1\endcsname[1]{{\bf \color{#2} \capitalisewords{#1}:\,##1}}
  \expandafter\newcommand\csname #1cor\endcsname[2]{{\color{#2} \capitalisewords{#1}:\,\st{##1}{\bf ##2}}}
  \expandafter\newcommand\csname #1color\endcsname{#2}
}
\addComment{cris}{blue} 
\addComment{james}{red}
\addComment{karthik}{cyan}
\addComment{patrick}{green}

\title{\boldmath Physics Event Classification Using Large Language Models}

\author{C. Fanelli, J. Giroux, P. Moran, H. Nayak, K. Suresh, E. Walter}
\affiliation{
William \& Mary, Williamsburg, VA 23185, USA}

\emailAdd{cfanelli@wm.edu}

\abstract{The 2023 AI4EIC hackathon was the culmination of the third annual AI4EIC workshop at The Catholic University of America. This workshop brought together researchers from physics, data science and computer science to discuss the latest developments in Artificial Intelligence (AI) and Machine Learning (ML) for the Electron Ion Collider (EIC), including applications for detectors, accelerators, and experimental control. The hackathon, held on the final day of the workshop, involved using a chatbot powered by a Large Language Model, ChatGPT-3.5, to train a binary classifier neutrons and photons in simulated data from the \textsc{GlueX} Barrel Calorimeter. In total, six teams of up to four participants from all over the world took part in this intense educational and research event. This article highlights the hackathon challenge, the resources and methodology used, and the results and insights gained from analyzing physics data using the most cutting-edge tools in AI/ML.}


\begin{document}
\maketitle
\flushbottom

\section{Introduction}
\label{sec:intro}

The AI4EIC2023 hackathon took place on the last day of the third annual AI4EIC Workshop from November 28 to December 1, 2023 at The Catholic University of America. This interdisciplinary workshop brought together researchers and students who were interested in and are working on AI/ML applications for Electron Ion Collider (EIC) physics.

The workshop included six sessions dedicated to specialized topics in EIC physics. The first session, \textit{AI/ML for ePIC and Beyond}, focused on AI/ML applications during the design phase of the ePIC detector as well as applications for the EIC science in general. The second session, \textit{Calibration, Monitoring, and Experimental Control in Streaming Environments}, featured talks on real-time data readout and analysis, and optimal control of detectors. The next session was \textit{AI / ML for accelerators}, including cutting-edge developments from leading accelerator facilities around the world. The fourth session, \textit{Foundation Models and Trends in Data Science with Tutorials}, presented some of the earliest research into the physics applications of foundation models, such as Large-Language Models (LLMs), one of the most booming subfields of artificial intelligence. The final session was \textit{AI in production, distributed ML}, discussing workflow management and computing architectures for AI/ML and particle physics.
The workshop also included introductory tutorials on specialized topics in ML such as Continual Learning and Reinforcement Learning. 

The entire final day of the workshop consisted of the AI4EIC hackathon, following on the success of the inaugural hackathon at the previous year's workshop \cite{allaire2024artificial}. The event involved both in person and remote participation of approximately two dozen researchers from institutions around the world. The main objectives of the hackathon were to better understand the capabilities of LLM, specifically ChatGPT, to address a machine learning (ML) challenge within the domain of experimental physics, followed by providing a rich introductory experience to participants to use LLM as a tool to better assist and increase their productivity.

This paper describes the structure and outcome of the hackathon. Section \hyperref[sec:problem_description]{\ref{sec:problem_description}} provides a description of the problem and relevant concepts of physics. Section \hyperref[sec:resources]{\ref{sec:resources}} provides details on the logistics and methodology of the hackathon, as well as the technical resources used. Finally, Section \hyperref[sec:conclusion]{\ref{sec:conclusion}} reflects on the insights gained from the hackathon and offers potential paths to continue this work.
The relevant code for the hackathon can be found in 
\href{https://github.com/ai4eic/AI4EICHackathon2023-Streamlit}{https://github.com/ai4eic/AI4EICHackathon2023-Streamlit}.

\section{The Hackathon Problem}
\label{sec:problem_description}

\paragraph{The LLM problem:}
Participants are tasked with developing a machine learning (ML) model for a classification problem of experimental physics using only an LLM (ChatGPT) interface. The scope of the hackathon is to evaluate the following. 
\vspace{-0.2cm}
\begin{enumerate}
    \setlength{\itemsep}{-2pt}
    \setlength{\parsep}{-2pt}
    \item \textit{LLM for Code Assist}: Creating the scripts necessary to build a machine learning model for binary classification with limited dataset information by using a Language Model (LLM) and making use of its responses.
    \item \textit{Few-Shot Prompting}: Constructing the best-performing ML model for binary classification task while minimizing the number of total number of prompts used.
\end{enumerate}
\vspace{-0.2cm}
To better align with the scope of the problem, participants had to be restricted to editing the scripts produced by ChatGPT and have limited knowledge of the dataset itself. Several unique rules were established, forming the foundation upon which the hackathon infrastructure was built, as follows:
\begin{itemize}
    \setlength{\itemsep}{-2pt}
    \setlength{\parsep}{-2pt}
    \item Participants could only use the provided custom Chat interface built for this hackathon to interact with ChatGPT3.5. This interface served as the primary platform for participants to formulate queries and receive responses from ChatGPT, allowing seamless interaction with LLM throughout the hackathon.
    \item Furthermore, participants did not have direct access to the datasets; this restricted participants to train the network elsewhere other than the provided infrastructure and could assess submissions on equal footings. 
    \item Participants did not have access to any editors or could edit the code snippets provided by ChatGPT3.5. Participants were even forced to use ChatGPT to install libraries to run code snippets. 
    \item Finally, to minimize the total number of prompts needed to build the model, all chat interactions were broken down into sessions. Each chat session had a fixed conversational history beyond which the chat session would restart. 
    \item It has been shown that ``in-context learning" improves LLM performance \cite{wang2024learning}. To facilitate this feature, participants were provided with the option to set the `session context' at the start of each chat session.
\end{itemize}
\vspace{-0.2cm}
The whole hackathon was divided into two questions with increasing difficulty to solve.

\begin{figure}[htbp]
\centering\includegraphics[height=7cm]{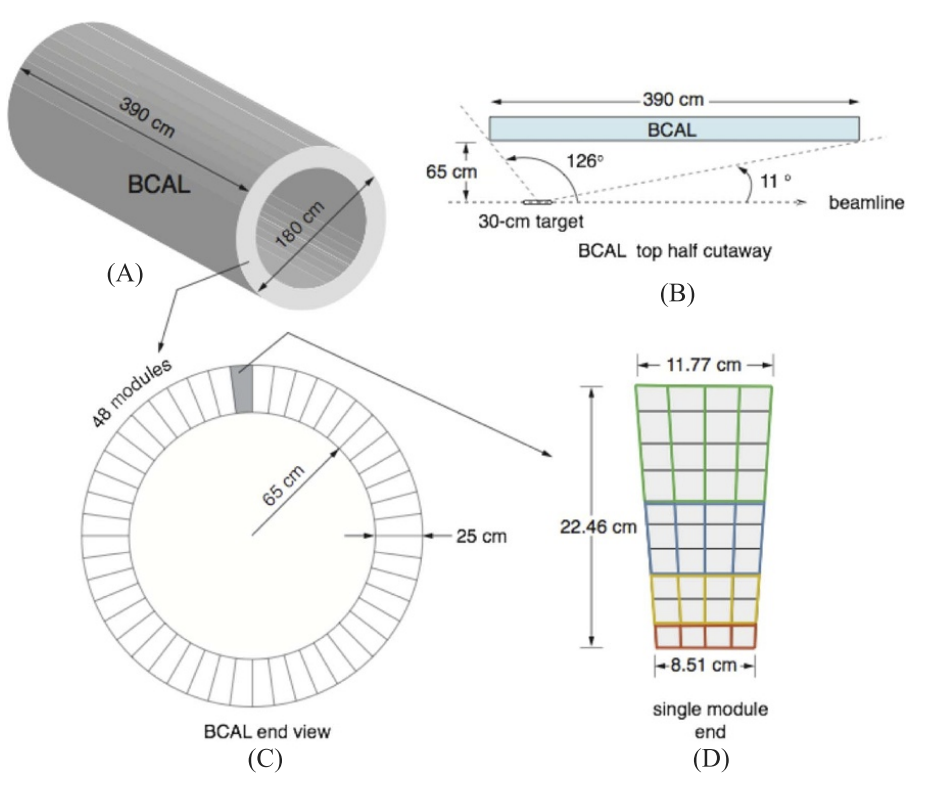} 
\caption{Sketch of barrel calorimeter readout: (A) BCAL schematic; (B) a BCAL module side view; (C) and view of the BCAL showing all 48 modules and (D) an end view of a single module showing readout segmentation in four rings (inner to outer) and 16 summed readout zones demarcated by colors. Figure from Fanelli et. al. ~\cite{F+M} \label{fig:gluex_readout}}
\end{figure}

\paragraph{The Classification Problem: PID with \textsc{GlueX} BCAL}
The \textsc{GlueX} experiment, one of the four major physics experiments in Jefferson Lab, aims to provide experimental validation of the theoretical predictions of Quantum Chromodynamics (QCD) \cite{gluex}. A crucial component of this experimental validation lies within the correct identification of final-state particles incident on the detector systems at \textsc{GlueX}. The hackathon primarily focuses on the classification of neutral shower candidates within the \textsc{GlueX} Barrel Calorimeter (BCAL), namely photons and neutrons. 
The BCAL is a 400 m long cylindrical sampling electromagnetic calorimeter (translating to a polar angle coverage of $\sim$11-$126^{\circ}$) \cite{gluex}. The BCAL is designed primarily for photon detection, with fine-grained energy resolutions ranging from $0.05$ to a few GeV. The calorimeter consists of layers of scintillating fibers interleaved with lead. 
As seen in Figure~\ref{fig:gluex_readout}, the BCAL consists of 48 azimuthal regions, each of which is divided into four readout layers, each consisting of double-ended readout through silicon photomultipliers (SiPM). 

The shower profile of neutral particles, \textit{e.g.}, photons, and neutrons, is directly dependent on the phase space, characterized by the reconstructed $z$ position of a shower and the reconstructed energy $E$ \cite{F+M}. 
In specific regions of the phase space, the interaction of neutrons within the calorimeter is limited, resulting in little energy deposition compared to photons and generally occurring in the outer layers. In contrast, photons tend to deposit the majority of their energy in the first few layers. In this case, the separation of the two classes is relatively simple and is achievable by rectangular cuts. However, in other regions neutrons tend to mimic photons to a high degree, making the separation of the two classes much more difficult. In this case, the usage of machine learning is optimal for providing non-linear separation.

\section{Infrastructure, Resources and Methodology}\label{sec:resources}

\vspace{-0.2cm}
\paragraph{Datasets.} Simulated data from Hall D was used for the hackathon challenge and made available to participants via their AWS instances. These archived data were generated using the GEANT4 particle gun to create high-purity neutron and photon samples in the \textsc{GlueX} BCAL. The samples utilized the \textit{recon-2019\_11-ver01.2}
reconstruction version and \textit{4.35.0} simulation version, and were constructed using the same conditions in Fanelli et al. \cite{F+M}, \textit{e.g.} a single shower condition to remove potential split-offs.\footnote{The term \textit{split-offs} refers to the conversion of a photon to an $e^+ e^-$ pair.}
The data were divided into two problems formulated as follows:
\vspace{-0.2cm}
\begin{itemize}
    \setlength{\itemsep}{-2pt}
    \setlength{\parsep}{-2pt}
    \item Problem 1 - Full phase space coverage. High degree of separability.
    \item Problem 2 - Reduced phase space coverage. Lower degree of separability.
\end{itemize}
\vspace{-0.2cm}
In the second problem, we specifically isolate a neutron population that mimics photons to a higher degree by imposing rectangular cuts on the radius, energy deposition in the outermost ($4^{th}$) layer, the fraction of energy deposited within the $2^{nd}$ layer, width in time, and width in $\phi$. Moreover, we purposely made the dataset imbalanced, heavily favoring the photon population.
Careful considerations were taken to ensure that the separation of the two classes was not ambiguous and that participants could still obtain encouraging results.

In each problem, separate training and test data sets were provided to participants. The data sets consisted of 14 feature variables describing the properties of electromagnetic showers \cite{F+M}. The definitions of the characteristic variables are given in Appendix~\ref{sec:data_defs} and include: the radial position of the shower, the energy deposited in each of the four BCAL layers, the energy fraction in each of the layers, and the widths of the shower in the z direction, radial direction, time dimension, azimuthal direction, and polar direction. 
Each row in the dataset represented an individual shower event originating from either a neutron or a photon. In addition, the training data included a column for the particle ID: 0 for neutrons and 1 for photons. The testing data included an additional column for the event ID, to allow for comparison with the answer key. The phase space of both neutrons and photons was chosen to represent a highly active region of the BCAL, specifically, $E = 200-2200$ MeV and $z = 162-262$ cm. 

\vspace{-0.2cm}
\paragraph{Time constraints.} The hackathon began at 10:00 EST on December 1, 2023 and ended at 17:00 EST on the same day. The hackathon was preceded by a tutorial at 09:00 EST that described the hackathon challenge to the participants.

\vspace{-0.2cm}
\paragraph{Context constraints.} The LLM was built as a ChatBot with the ability to retain conversational history. Each conversation therefore is referred to as sessions. Participants can stop and start a fresh session. However, the length of the conversation was limited. This mainly involves evaluating the ability to initiate a fully functional and efficient model to solve the problem. The total conversational length was constrained to 12000 tokens. This translates to about 8000 characters \cite{openai-gpt3.5}. If the user conversation exceeds the maximum number of tokens, then the session automatically resets to a new session. 

\begin{figure}
    \centering
    \begin{subfigure}[t]{0.475\textwidth}
        \includegraphics[scale = 0.225]{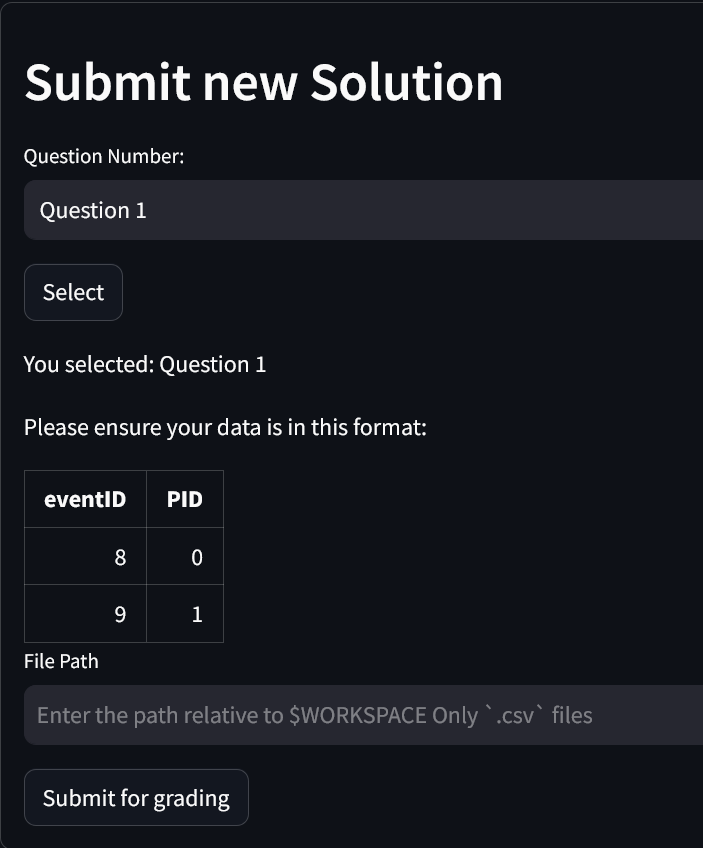}
        \subcaption{Submission portal}
        \label{subfig:submission}
    \end{subfigure}
    \begin{subfigure}[t]{0.475\textwidth}
        \includegraphics[scale = 0.1]{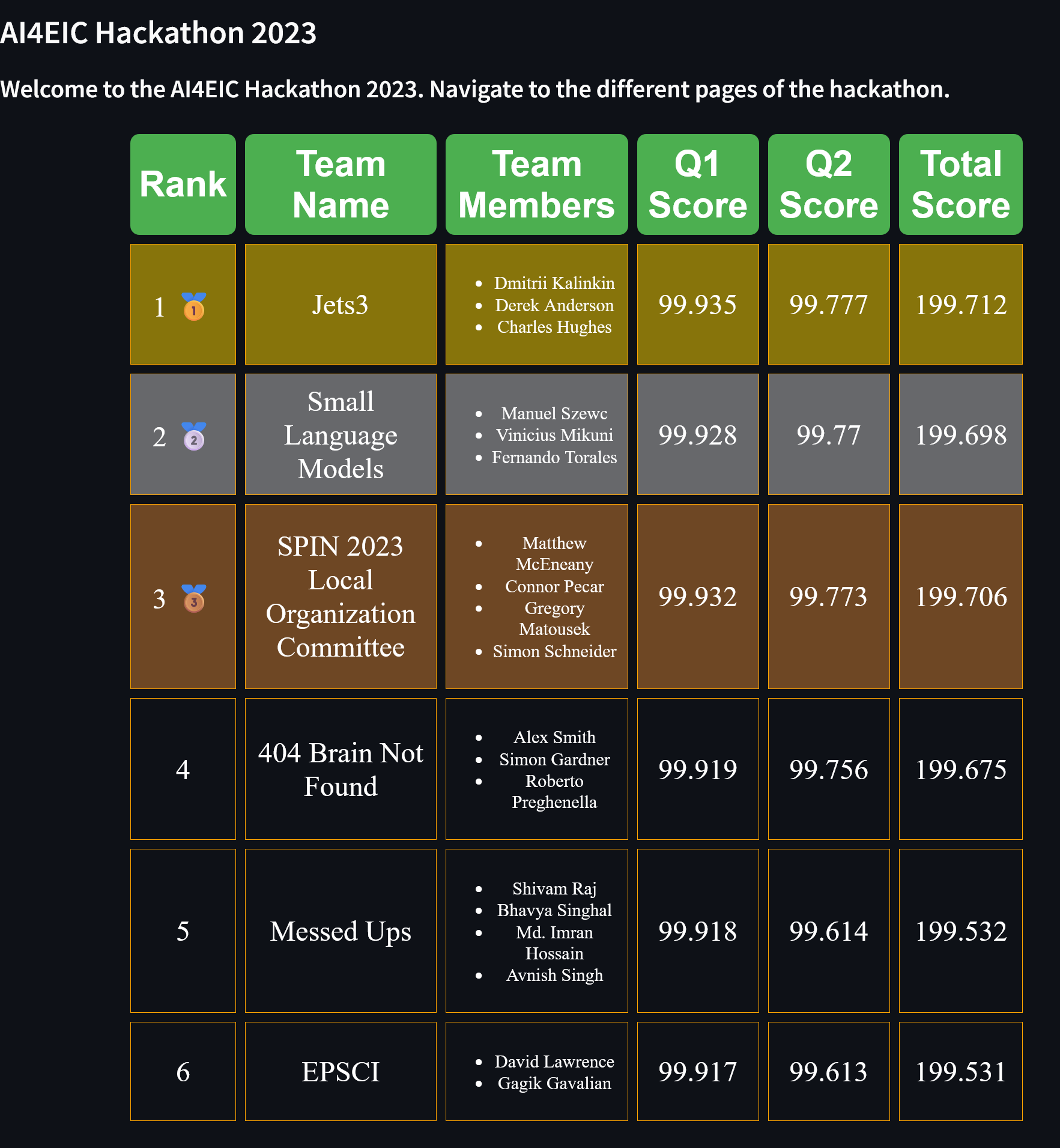}
        \subcaption{Final Leaderboard}
    \end{subfigure}
    \caption{(a) The plot on the left, shows the interface, provided for submission of solutions. Users select the question to evaluate and the respective file path in the AWS instance. Once submitted, the solutions are graded and automatically update in the leaderboard. (b) The figure on the right shows the leaderboard at the end of the hackathon summarizing the performance stats of the teams. \uline{With LLM, all participants achieved an accuracy of more than $99\%$ for both questions and are within the statistical fluctuations}. However, ``Jets" used the least number of tokens (8192 tokens in total) for the submitted solutions and were the fastest compared to other participants. }
    \label{fig:Final-leaderboard}
\end{figure}

\vspace{-0.2cm}
\paragraph{Scoring.} The classified events in the \texttt{submitted.csv} files were compared with the ground truth events from the simulation. The score for each question was taken to be the percentage of events that are correctly classified, that is, the precision of the classifier. Participants were instructed not to change the event identification number of the test set, as this would result in an essentially random comparison of events.
%
%
%
The overall score was the sum of the scores for Question 1 and Question 2. Hence, the highest possible score for each question was 100, and the highest possible overall score was 200.
The uncertainty in the metric was calculated using Eq.~\ref{eq:uncertainty}:


\begin{equation}\label{eq:uncertainty}
    \sigma = \sqrt{\frac{Accuracy \cdot (1 - \text{Accuracy})}{N}},
\end{equation}
\smallskip
where $N$ is the number of events to label.
%
Participants were instructed to submit their results through the submission portal on the web application, in which an example of the \texttt{required.csv} format was provided. The submission interface can be found in Fig.~\ref{fig:Final-leaderboard}. 
Two additional metrics are considered to rank the teams if there is a tie:
\vspace{-0.2cm}
\begin{itemize}
    \setlength{\itemsep}{-2pt}
    \setlength{\parsep}{-2pt}
    \item If the scores of the teams are within statistical fluctuations. The team with the minimum number of prompts taken to achieve the best performing solution will precede in the ranking.
    \item If the tie continues, the team that submitted the best solution first precedes in the ranking.
\end{itemize}

\subsection*{Software Infrastructure and Design}
\begin{figure}
    \centering
    \includegraphics[scale = 0.425]{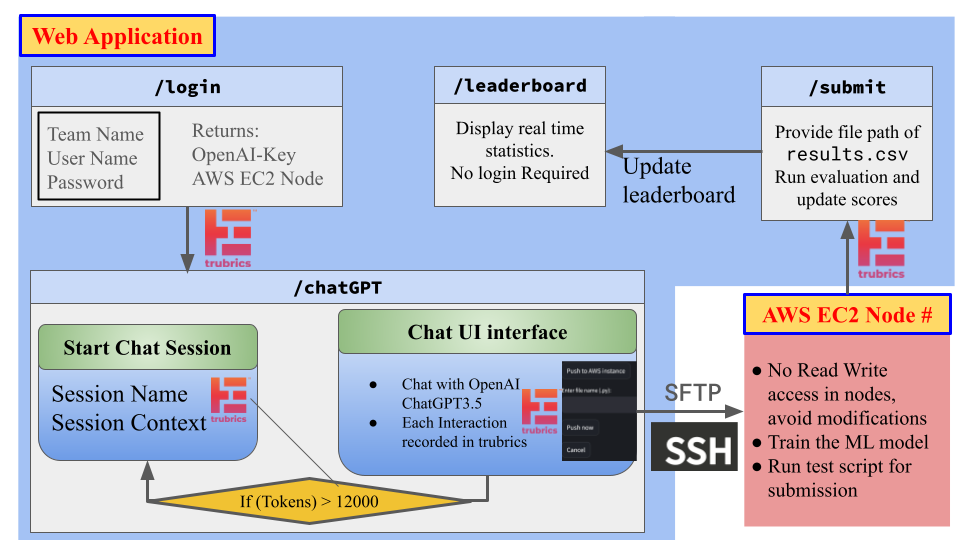}
    \caption{The developed software infrastructure for the hackathon. 
     The arrows in the figure denote the flow of control for the application. 
    Interactions between users and ChatGPT are recorded through \texttt{Trubrics}.
    A sample Chat Session is shown in the appendix in Fig.~\ref{app:example_prompt}.}
    \label{fig:software-infrastructure}
\end{figure}

The software infrastructure designed for this hackathon comprises of two distinct tasks. The initial task focuses on developing a web application that allows users to interact with ChatGPT3.5 via a chat interface to create suitable scripts for building the machine learning model. The subsequent task revolves around securely transferring these scripts to compute resources where the data sets are available for training. Interaction between the compute nodes and the web server is necessary for script transfer and result retrieval during the evaluation process. Fig. \ref{fig:software-infrastructure} shows the software infrastructure developed for the hackathon. Detailed information on the two tasks is summarized in the following.

\vspace{-0.2cm}
\paragraph{Compute Resources.}
Users were instructed to form teams of up to four participants, in which each team member received their own AWS \textit{g5.2xlarge} instance with a single Nvidia A10G GPU, 8 cores, 32 GB of RAM and 450 GB of disk \cite{aws-resource}. The choice of instance was based on problem complexity, and with the idea of minimizing training times users may incur by providing adequate GPU support. Each instance was pre-built with conda, in which users were instructed to create their own environments based on the packages they intended to use. While not explicitly stated, an efficient method of creating conda environments was directly through ChatGPT, in which accurate bash scripts could be provided given informative prompts. Various conda test environments used by the organizers were constructed using this methodology.

\vspace{-0.2cm}
\paragraph{Web Application.}
The user interface was built using the \texttt{Streamlit} \cite{streamlit} platform, specifically built for LLM interfaces. The application was hosted on an AWS \textit{ c5.24xlarge}, where users could access the website using the provided URL. The instance consisted of 96 cores and 192 GB of RAM, allowing for efficient distribution of GPT inquiries asynchronously. Each user was issued log-in credentials, which allowed full access to the Web application, including access to ChatGPT 3.5 Turbo \cite{openai-gpt3.5} with a default context related to physics, as shown in \ref{fig:software-infrastructure}. 
Users were able to change the context of their AI assistant at the start of any new chat session. This also allowed efficient data collection on a per-user basis. Specifically, we were able to record interactions between users and ChatGPT through the \texttt{Trubrics} \cite{trubrics} collection API in the form of prompts and LLM responses. Each user account was directly connected to their AWS instance provided in such a way that a seamless transition of the GPT-provided code could be securely pushed to their instance with the click of a button. Users could also ask GPT to provide bash scripts to create environments and install the packages necessary for their analysis. Appendix~\ref{app:example_prompt} contains an example prompt received during the hackathon and the corresponding output.

\vspace{-0.2cm}
\paragraph{Results: Participants Exceeding Expectations}
The participants and their AI-assisted coding strategies \uline{greatly exceeded our expectations in the second problem}. Internal tests prior to the hackathon indicated an expected accuracy of around $92\%$, under the same constraints as the participants, but with prior knowledge of the data set. We requested ChatGPT to provide us with a syntactically correct program, deploying an under-sampling technique to combat imbalanced data, a basic XGBoost \cite{Chen_2016} decision tree, and a performance analysis as a function of decision boundary placement in the validation set to find the optimal threshold of the model. \uline{Remarkably, all teams and participants managed to greatly exceed this score with unique solutions}. In fact, all teams submitted scores approaching $\geq 99 \% $ and are within statistical error. The final decision came down to the combination of the minimum number of prompts to obtain their result and the time submitted in which the ``Jets' team emerged victorious. Their strategy involved a CatBoostClassifier \cite{prokhorenkova2018catboost} coupled with a hyperparameter optimization technique, all of which were directly accessed via concise commands to ChatGPT.

\section{Conclusions}\label{sec:conclusion}

With the emergence of powerful predictive text tools such as ChatGPT, it is important to understand the context in which these tools should and can be applied. 
The hackathon extended its scope beyond problem-solving to encompass educational outreach initiatives targeted at researchers, scientists and practitioners within the Nuclear and Particle Physics communities. These efforts have created awareness of the capabilities of LLM as an effective tool to better assist and increase productivity. LLM is also capable of providing detailed code explanations, which is extremely beneficial for those new to the field. 
In fact, we received comments in which the LLM was cited as a direct reason for the completion of the ML task given to participants. The saved metadata amounted to a total of 752 prompts between 19 unique users during the 8-hour period of the hackathon. 

\vspace{-0.2cm}
\paragraph{Future studies based on Hackathon:} The hackathon also served as a test bed for data collection to further study the capabilities of LLM on Few-Shot Prompting \cite{brown2020language} and eventually Zero-Shot Prompting \cite{wei2022finetuned} for a specific domain task such as the hackathon problem. All the prompts and responses for each of the users, along with additional metadata, were collected during the course of the hackathon. Most of the participants in the hackathon were domain experts. Therefore, analyzing the quality and precision of the responses generated by ChatGPT for domain experts-engineered prompts could lead to the development of few-shot prompt methods for this specific ML task. This helps to identify strategies to leverage domain expertise to enhance the quality and relevance of prompts, such as providing additional context, specifying problem constraints, or suggesting relevant features for classification of such tasks.
Other studies that could be done on the collected data along with more data collected from ``non-domain" experts are summarized below.
\vspace{-0.2cm}
\begin{itemize}
    \setlength{\itemsep}{-2pt}
    \setlength{\parsep}{-2pt}
    \item \textbf{Performance Analysis}: Measure performance metrics such as response completion time, code efficiency from the participants, and evaluate it against the responses generated by the prompts from non-domain experts.
    \item \textbf{Error Analysis}: Determine whether prompts engineered by domain experts lead to fewer errors, more accurate solutions, or better alignment with the problem requirements compared to those from nonexperts.
\end{itemize}


\acknowledgments

The authors acknowledge the Catholic University of America for hosting the AI4EIC workshop's hackathon. We also thank the College of William \& Mary for their support, and Research Computing for providing computational resources and/or technical support that have contributed to the results reported in this article (\href{https://www.wm.edu/it/rc}{https://www.wm.edu/it/rc}). The authors thank AWS for providing the compute resources for this workshop.  
The authors acknowledge the \textsc{GlueX} \cite{GlueX-Collaboration} collaboration for allowing us to use the simulated MC event samples as data sets for the problems.


\bibliographystyle{JHEP}
\bibliography{biblio.bib}

\appendix

\section{Definitions of Feature Variables}
\label{sec:data_defs}

   The overall features of the shower are indicated by a subscript S. For example, $S_{z}$ represents the shower z-position. $T_{z}$ denotes the z-position of the center of the target, and R denotes the inner radius of the BCAL. Most of the variables represent energy-weighted sums of individual hits, represented with the summation subscript i.

    \begin{itemize}
        \item \textbf{r} = $\frac{1}{E_{tot}}\sum_{i}^{N} r_{i} E_{i}$ (i.e. the energy-weighted radial position, in cm)
        \item \textbf{LayerM\_E} = The energy deposited in the Mth layer of the BCAL, in MeV ($M \in \{1,2,3,4\}$).
        \item \textbf{LayerMbySumLayers\_E} = The fraction of the total energy deposited in layer M of the BCAL ($M \in \{1,2,3,4\}$).
        \item \textbf{ZWidth} = $\sqrt{\frac{1}{E_{tot}}\sum^{N}_{i}E_{i} (\Delta z_{i})^{2}}, \, \Delta z_{i}=(z_{i}+T_{z})-S_{z}$ (ie, the energy-weighted shower width in the z-direction, in cm).
        \item \textbf{RWidth} = $\sqrt{\frac{1}{E_{tot}}\sum^{N}_{i}E_{i} (\Delta r_{i})^{2}}, \, \Delta r_{i}=R-r_{i}$ (that is, the shower width weighed on energy in the radial direction, in cm).
        \item \textbf{TWidth} = $\sqrt{\frac{1}{E_{tot}}\sum^{N}_{i}E_{i} (\Delta t_{i})^{2}}, \, \Delta t_{i}=t_{i}-S_{t}$ (that is, the temporal width of the shower, weighted by energy, in ns).
        \item \textbf{PhiWidth} = $\sqrt{\frac{1}{E_{tot}}\sum^{N}_{i}E_{i} (\Delta \theta_{i})^{2}}, \, \Delta \theta_{i}=\theta_{i}-S_{\theta}$ (that is, the shower width weighted with energy in the azimuthal direction, in radians).
        \item \textbf{ThetaWidth} = $\sqrt{\frac{1}{E_{tot}}\sum^{N}_{i}E_{i} (\Delta \phi_{i})^{2}}, \, \Delta \phi_{i}=\phi_{i}-S_{\phi}$ (i.e., the energy-weighted shower width in the polar direction, in radians).

    \end{itemize}

\section{The Secret Context}
\label{sec:secret_context}
A system prompt is a way to provide context and instructions to ChatGPT, such as specifying a particular goal or role for LLM before asking it a question or giving him a task. Therefore, a prompt was given specifying the behavior of the ChatGPT interface. This is to avoid any deception. 
The secret prompt that is set is given below.
\vspace{-0.2cm}
\begin{quote}
    \textit{
    You are an expert python programmer, very proficient in the following python packages. \\
        1. \texttt{numpy} \\
        2. \texttt{pandas} \\
        3. \texttt{pytorch} especially using cuda for GPU acceleration \\
        4. \texttt{hdf5} \\
        5. \texttt{tensorflow} \\
    You are very critical in writing code with no run-time errors. You can write code snippets in Python. \\
    You will strictly not answer questions that are not related to programming, computer science, and hardonic physics. \\
    Politely decline to answer any conversation that is not related to the topic. \\
    }
\end{quote}

\section{Example Prompt and Output}
\label{app:example_prompt}

\begin{figure}[!htbp]
    \centering
    \includegraphics[scale = 0.15]{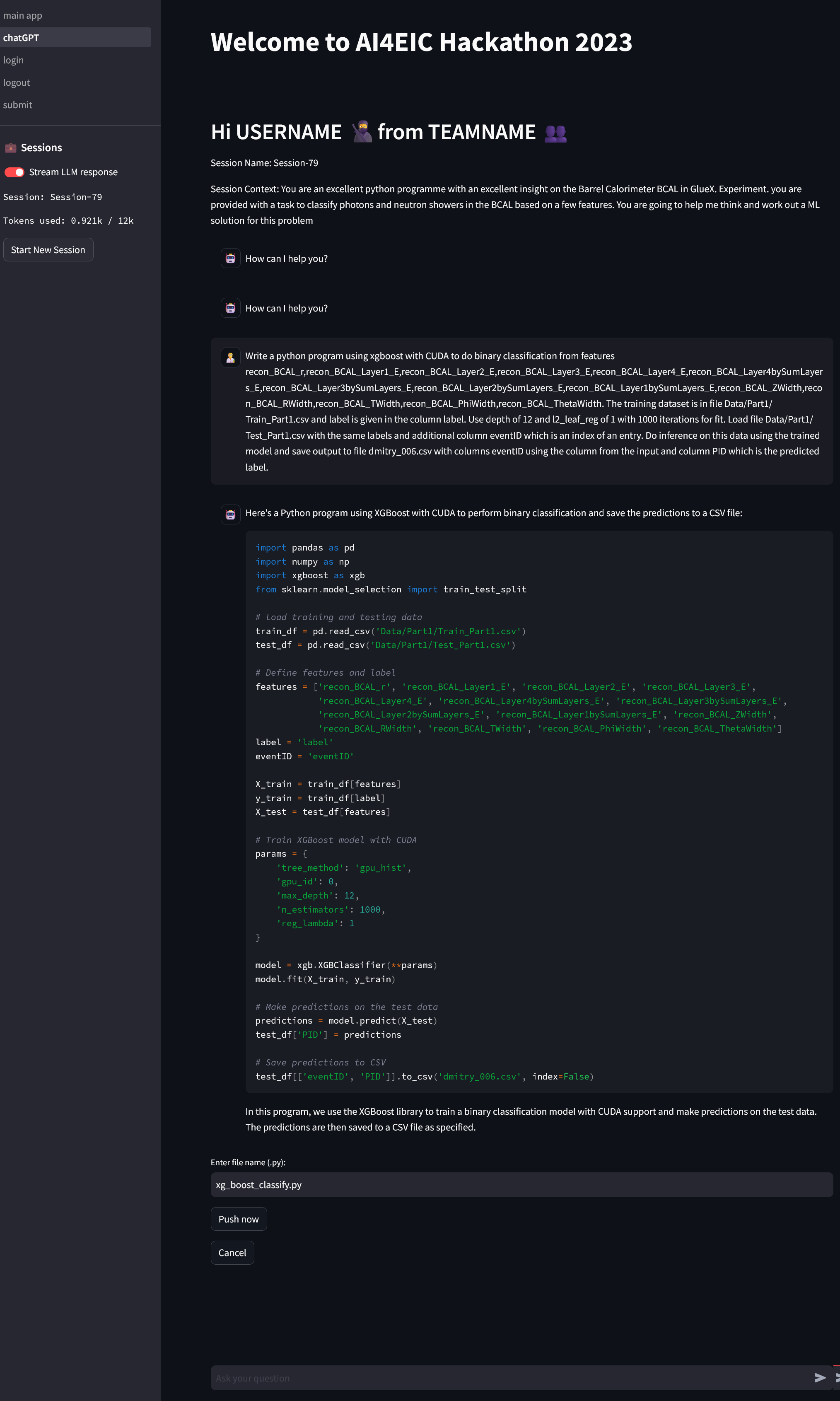}
    \caption{\textbf{Example GPT Chat Session:} 
    In a hackathon, a participant starts a chat session by setting a context, then asks ChatGPT questions. ChatGPT provides code and explanations, which the participant names and pushes to AWS for training. The session ends upon code submission, redirecting the user to a portal. Usage statistics are displayed, indicating when the session will end and a new one will start based on token usage.
    }
    
    \label{fig:example_prompts}
\end{figure}

\end{document}